\documentstyle[epsfig]{aa}    

\newcommand{\Msun} {M$_\odot$}

\begin{document}
 
\thesaurus{08.03.4, 08.16.5, 09.12.1   }
 
\title{Mid infrared emission of nearby Herbig Ae/Be
stars\thanks{Based on observations with ISO, an ESA project with
instruments funded by ESA Member States (especially the PI countries:
France, Germany, the Netherlands and the United Kingdom) with the
participation of ISAS and NASA.}}
 
\author { R.~Siebenmorgen\inst{1}, T.Prusti\inst{2}, 
                A.  Natta\inst{3}, T.G. M\"uller\inst{2}}
 
\institute{
 \inst{2} European Southern Observatory,
	 Karl-Schwarzschildstr. 2,
	 D-85748 Garching b. M\"unchen, Germany \\
 \inst{2} ISO Data Centre,
                Astrophysics Division, ESA,
                 Villafranca del Castillo,  P.O. Box 50727, 
		 E-28080 Madrid, Spain \\
 \inst{3} Osservatorio Astrofisico di Arcetri,
                Largo E. Fermi 5, I-50125 Firenze, Italy}
 
\offprints{rsiebenm@eso.org}
\date{Received 8 May 2000; accepted 21 June 2000 }
\authorrunning {Siebenmorgen et al.}
\titlerunning {The evolution of mid infrared emission of Herbig Ae/Be
stars}
\maketitle
 
\begin{abstract}
 
We present mid IR spectro-photometric imaging of a sample of eight
nearby ($D \leq 240$pc) Herbig Ae/Be stars.  The spectra are dominated
by photospheric emission (HR6000), featureless infrared excess
emission (T~Cha), broad silicate emission feature (HR5999) and the
infrared emission bands (HD~97048, HD~97300, TY~CrA, HD~176386). The
spectrum of HD179218 shows both silicate emission and infrared
emission bands (IEB). All stars of our sample where the spectrum is
entirely dominated by IEB have an extended emission on scales of a few
thousand AU ($\sim 10''$). We verify the derived source extension
found with ISOCAM by multi--aperture photometry with ISOPHT and
compare our ISOCAM spectral photometry with ISOSWS spectra.

\end{abstract}
 
\section{Introduction}

The mid-infrared spectrum of Herbig Ae/Be stars is rich in information
that can be used to improve our understanding of the circumstellar
environment. The wavelength interval between 6 and 13 $\mu$m contains
most of the infrared emission bands (IEB), as well as the broad
silicate feature centered at about 9.7 $\mu$m.

From ground-based and KAO spectroscopy (Roche et al. 1991, Schutte et
al. 1990) it is known that IEB exist in the spectra of some Herbig
Ae/Be stars.  A large number of Herbig Ae/Be stars have been measured
with the Infrared Space Observatory (ISO). The ISO sample will allow
us to determine the properties and nature of the IEB carriers with
unprecedented accuracy (Waelkens et al. 1996).  These data will be
crucial in solving the puzzle of the mid-infrared energetics of Herbig
Ae/Be stars, by determining to what extent IEB emission contributes to
the observed mid-infrared luminosity, and how much IR emission comes
from a circumstellar disk in Herbig stars of different spectral type
and age (Kenyon \& Hartmann 1991, Hillenbrand et al. 1992, Natta et
al. 1993).

A very important contribution to our understanding of the IEB and of
their role in Herbig Ae/Be stars is provided by spatial information on
the extension and distribution of the emission at various
wavelengths. For example, Prusti et al. (1994) were able to infer the
presence of infrared emission bands (IEB) and to determine some of the
properties of the distribution of the carriers around Herbig Ae/Be
stars from multi--aperture photometry in the three narrow-band filters
N1, N2 and N3.  Natta \& Kr\"ugel (1995), using the radiation transfer
code developed by Siebenmorgen et al. (1992), computed the expected
intensity profile in the IEB and in the adjacent continuum as a
function of the spectral type of the exciting star and of the
surrounding shell parameters.  However, until recently the scarcity of
observed intensity profiles has greatly limited the usefulness of
these calculations.

Spectro-photometric imaging observations using the circular variable
filters (CVF) of ISOCAM (Cesarsky et al. 1996) are capable of
producing the spatial information we want. In addition to the
determination of the source size at the various wavelengths, they can
also reveal additional emission structures and deviations from
spherical symmetry. We present in this paper the results of ISOCAM
spectro--photometric, ISOPHT (Lemke et al. 1996) multi--aperture and
ISOSWS (de Graauw et al. 1996) grating scan observations of a small
sample of Herbig Ae/Be stars that we consider particularly suited to
this purpose.

Our sample includes 5 stars of spectral type B9/A0 (HD~179218,
HD~176386, TY~CrA, HD~97048 and HD~97300). They cover a small age
interval, from 1 Myr to $>3$ Myr (Table 1).  Some of these stars
(i.e., TY~CrA and HD~97048) are known to have strong IEB features
(Roche et al. 1991, Deutsch et al. 1995). In addition, Prusti et
al. (1994) could infer from multi--aperture study the presence of IEB
for HD176393, HD97048 and HD~97300.

In addition to these 5 stars, we have included in our sample two stars
of later spectral type, namely HR~5999 (spectral type A5/7) and T~Cha
(spectral type G8). The aim was to check the dependence of IEB band
shapes and intensity profiles on the spectral type of the exciting
star. Ground-based observations in the 10 $\mu$m region (Wooden 1994)
seem to indicate that, as expected, IEB are weak or absent in Herbig
Ae/Be stars of later spectral type (see also the 3.3 $\mu$m survey of
Brooke et al. 1993).  Much more surprisingly, it was also found that
the silicate feature appears in emission only in stars of later
spectral types. There is no obvious reason why silicate emission
should disappear in hotter stars.  Its absence is probably due to the
geometrical distribution of the dust around the star. If we believe
that the silicate emission is linked to the existence of a
circumstellar disk (Chiang \& Goldreich 1997), it is possible that
early-type stars do not show the silicate feature because they do not
have disks.  This would confirm the results obtained with millimeter
interferometry, namely that Herbig stars of spectral type A show
evidence of disks while very few Herbig B stars do (Di Francesco et
al. 1994).  It is possible that silicate emission disappears in B
stars as they age, dissipating their disks and pushing farther away
the residual envelopes. The discovery of "transitional" objects,
showing both IEB and silicate emission would be very interesting in
this context.

A convincing answer to these questions requires a statistical (i.e.
large) sample of spectra and intensity maps for Herbig Ae/Be stars of
different spectral type and age.  However, we feel that our results
can provide interesting clues to a possible answer.

This paper contains in Sect.~2 a description of the observations and data
reduction techniques. The results for the individual stars are
presented and discussed in Sect.~3. A summary is given in Sect.~4.

\section{Observations}

Spectro-photometric images have been obtained with ISOCAM for the
stars in our sample in a number of CVF wavelength intervals, selected
to probe the emission in IEB at 6.2, ''7.7'', 8.6 and 11.3\,$\mu$m.
Additionally, one interval was chosen around the peak of the silicate
feature at 9.7\,$\mu$m. Typically, we have taken 25 exposures of
2.1\,s at each CVF wavelength. We used a gain of 2 and the lens
providing 3$''$ pixel field of view. The basic data reduction steps
are described in Siebenmorgen et al. (1999). After dark current
subtraction, glitch removal and flux transient correction, the
exposures are coadded in each CVF step. Photometry is obtained
applying the spectral response function. After background subtraction
we simulate multi--aperture photometry on the images.

\begin{figure}[!ht]
  \begin{center}
    \leavevmode
  \centerline{\epsfig{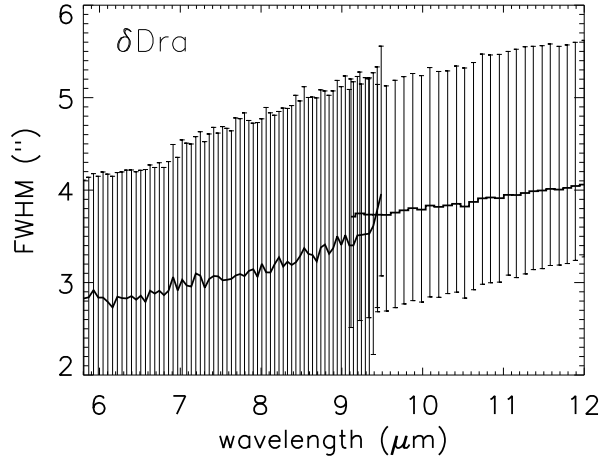}}
  \vspace{-0.5cm}   \end{center}
%
%
\caption{The size (FWHM) of the comparison star $\delta$~Dra as
derived from multi--aperture photometry on the CVF images. }
\label{fig:dDra}
\end{figure}

In order to understand the photometric uncertainty of this procedure
we analyzed data of the calibration star $\delta$~Draconis and
compared it with the Kurucz model.  The largest uncertainty was found
in the overlap region of CVF1 and CVF2 and is typically of the order
of $\le 5\%$.  The aperture sequence allows us to determine the full
width at half maximum flux (FWHM). For a point source
($\delta$~Draconis, as shown in Fig.~\ref{fig:dDra}), we derive in the
wavelength range considered, typical sizes of 3--4$''$ (FWHM) with an
uncertainty of about half a pixel size ($\sim 1.5''$).
 
We searched in the ISO archive and whenever possible we compared our
ISOCAM spectral photometry with ISOSWS data. The ISOSWS spectra are
from the off--line processing version 8.6 (Leech et al. 2000).

The CVF images are contaminated by ghosts and stray light. In order to
independently check the FWHM as derived by ISOCAM we compare them to
results obtained by ISOPHT.  We selected multi--aperture photometry
(PHT04, Laureijs et al. 2000) in 11 apertures ranging from 5$''$ up
to 180$''$ at an integration time of 32s per filter.  The dark
subtracted, deglitched and linearised signals of our target stars are
compared to an identical PHT04 sequence measured on a point source
(HR7127). Such a direct comparison is necessary because the point
sources do not exactly follow the theoretical PSF behaviour (M\"uller
2000). However, adequate point source sequences exist only in a
limited number of filters, in practice we can present our PHT04
measurements only at 7.3$\mu$m.

\begin{table*}
\begin{center}
\caption{\em Our ISO sample of Herbig Ae/Be stars}
\leavevmode
\begin{tabular}[h]{llcrrcl}
\hline \\[-5pt]
 Name	& ST        & Distance	  & T   & L  	     & Age & Cloud   \\
        &           & (pc)        & (K) & (L$_\odot$)&(Myr) &  
       \\[+5pt]
(1)&(2)&(3)&(4)&(5)&(6)&(7)\\
\hline \\[-5pt]
HD~97048& A0        & 180$^{+30}_{-20}$  &10000 &41&$>$2& Chamaeleon~I \\
HD~97300& B9        & 188$^{+43}_{-30}$  &10715 &37&$>$3& Chamaeleon~I     \\
HD~179218& B9       & 244$^{+70}_{-44}$ &10715 &316& 0.1& no apparent cloud   \\
HD~176386& B9IV     & 140$^{+30}_{-20}$ &10715 &49& 2  & Corona Australis \\
TY~CrA  & B9        & 140 $^{+30}_{-20}$ & 10715&98& 1  & Corona Australis\\
HR~5999	& A5-7III/IV&  210$^{+50}_{-30}$ &7943 &85& 0.5& Lupus 3 	  \\
HR~6000	& B6 V      &  241$^{+60}_{-41}$ &12882 &263& $>$0.5 & Lupus 3 \\
T~Cha   & G8        &  66$^{+19}_{-12}$   &5888 &1.3&$>$13& Chamaeleon~I/II   \\

\hline \\
\end{tabular}
\label{tab:table}
\end{center}
\end{table*}

\section {Results for  Individual Stars}

The stars in our sample are listed in Table 1.  Column 1 gives the
name of the star, Column 2 the spectral type, Column 3 the distance,
Column 4 the star effective temperature, Column 5 the luminosity,
Column 6 the age and in Column 7 we note the molecular cloud to which
the star belongs. Distances, effective temperatures, luminosities and
ages are taken from van den Ancker et al. (1996).  For TY~CrA, for
which no Hipparcos distance is available, we have taken the same
distance as of HD~176386 and computed the luminosity from the observed
V, (B-V) (Shevchenko et al. 1993), corrected for extinction with the
assumption that the ratio of selective-to-total extinction is {\it
R}=5.1 (Bibo et al. 1992). The age of the star is then derived by
comparing its location on the HR diagram to the evolutionary tracks of
Palla \& Stahler (1993).  The same procedure has been used to estimate
the luminosity and age of HR~6000.  This procedure is similar to that
used by van den Ancker et al. (1996).

\subsection {HD~97048 and HD~97300}

These two stars are located in the Chamaeleon~I cloud, and have very
similar properties. Their spectral type is A0 (HD~97048) and B9
(HD~97300); they have luminosities of 40 and 37 L$_\odot$,
respectively and lie very close to the ZAMS. Both stars have been
detected by IRAS. HD~97048 has a near-infrared excess, while HD~97300
has not.

\begin{figure}[!ht]
  \begin{center} \leavevmode
    \centerline{\epsfig{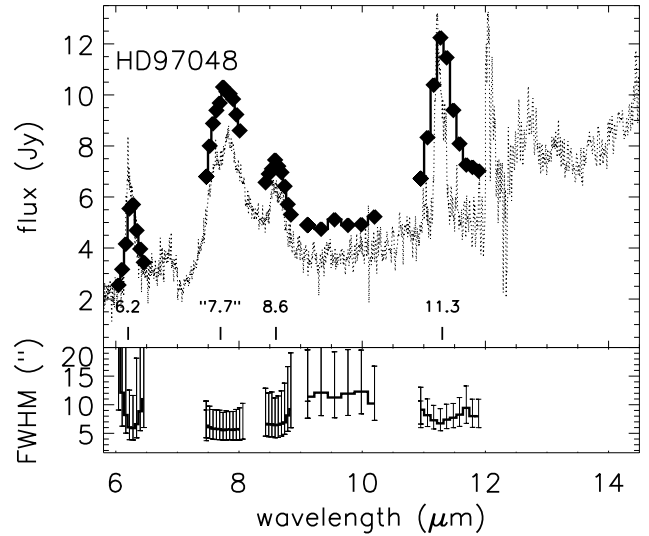}}
    \vspace{-0.5cm} \end{center} 
\caption{Top: Mid infrared spectrum of HD97048. The ISOCAM CVF
spectrum is indicated by $\Diamond$ and connected by the solid 
line.  The ISOSWS spectrum is shown by the dotted line. Bottom: The FWHM as
derived from multi--aperture photometry on the CVF images.}

\label{fig:HD97048_data}
\end{figure}

The mid-infrared spectrum of HD~97048 is known to be rich in IEB from
ground-based and KAO observations (Roche et al. 1991, Schutte et
al. 1990; Brooke et al. 1993). Prusti et al. (1994) found that the
emission was extended in both sources (by comparing aperture
photometry between 5$''$ to 16$''$), but more so in HD~97300 than in
HD~97048. Our ISOCAM and a ISOSWS spectrum of HD~97048 is shown in
Fig.~\ref{fig:HD97048_data}.  It is indeed dominated by IEB emission
centered at 6.2, ''7.7'', 8.6 and 11.3$\mu$m.  HD~97048 is extended on
scales of a few arcseconds (5--10$''$). Consequently the ISOSWS
spectrum (Van Kerckhoven et al. 1999), which is flux calibrated
assuming a point source, underestimates the total emission; its high
spectral resolution shows that the ''7.7'' $\mu$m band is due to two
separate components, which are unresolved by ISOCAM.

\begin{figure}[!ht]
  \begin{center} \leavevmode
    \centerline{\epsfig{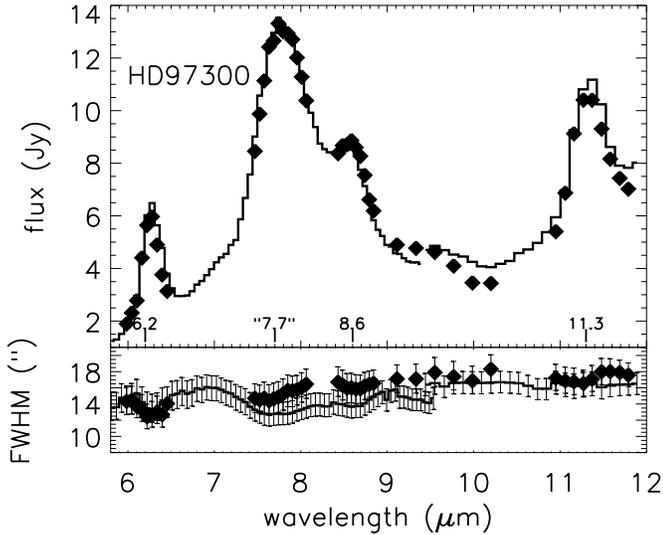}}
    \vspace{-0.5cm} \end{center} 

\caption{Top: Mid infrared spectrum of HD97300. The CVF full scan by
Siebenmorgen et al. (1998) is shown as solid line and the CVF sparse
scan of this paper by ($\Diamond$). Bottom: The FWHM as derived from
multi-aperture photometry on the CVF full scan (line) and sparse scan
($\Diamond$).}

\label{fig:HD97300_data}
\end{figure}

A deeper full CVF scan combined with photometric mid IR images of
HD97300 has been presented by us in a separate paper (Siebenmorgen et
al. 1998). It was found that the complete mid IR spectrum of the
source is dominated by a huge ring structure of about 7500 AU in
size. It was proven, by means of a dust model including transiently
heated particles, that all the ring emission is due to the IEB
carriers.  The amount of circumstellar matter was found to be rather
small ($\sim 0.03$\,\Msun).  In Fig.~\ref{fig:HD97300_data} we compare
the full CVF wavelength spectrum with the observing strategy of sparse
CVF scans, presented in this paper.  One notices that the absolute
photometry of both observations are identical down to a few \%.  Good
agreement is also found for the derived FWHM of both CVF scans. We
derive a FWHM of 13--19$''$, which gives a clear sign of a huge
extended nebulae as was independently verified by stray light free,
narrow band ISOCAM images.

\begin{figure}[!ht]
  \begin{center} \leavevmode
    \centerline{\epsfig{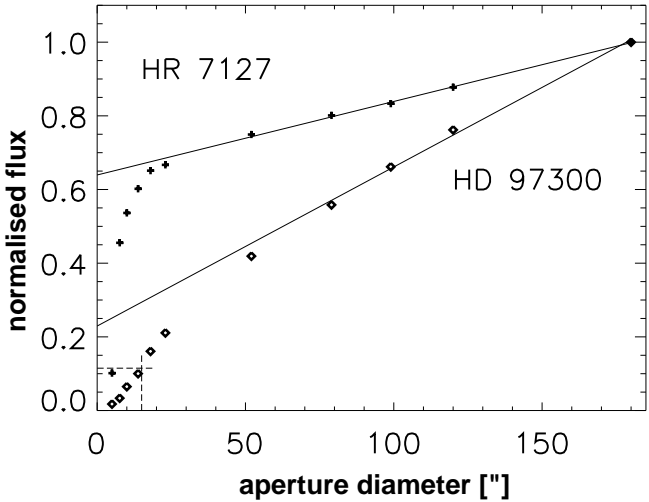}}
    \vspace{-0.5cm} \end{center} \caption{ISOPHT multi--aperture sequence
    on a point source (HR 7127) and the extended source HD 97300 at
    $7.3\,\mu$m. The line indicate the background estimate and the intersection at 0$''$ gives the source flux. The dashed line shows for HD~97300 the distance at half the source flux (FWHM).  The FWHM of HD~97300 as determined from this figure is $14 \pm 3''$.}
\label{fig:HD97300_pht}
\end{figure}

The FWHM of 14 $\pm 3''$ as derived from our ISOPHT multi-aperture
sequence at 7.3$\mu$m, presented in Fig.~\ref{fig:HD97300_pht}, is
consistent with the CVF results.

\subsection{TY CrA and HD\,176386}

The triple system TY~CrA (spectral type B9 of the primary) and the
binary HD\,176386 (spectral type B9 of the primary) are located in the
Corona Australis molecular cloud. The stars are separated from each
other by about 55$''$.  Both stars are located very close to the ZAMS.
Both have been detected by IRAS, but show no near-infrared excess, if
anomalous extinction is taken into account (Bibo et al. 1992). TY~CrA
has been detected at 1.3 mm, but the amount of circumstellar matter
associated to it is quite small (Hillenbrand et al. 1992; Natta et
al. 1997).  The mid-infrared spectrum of TY~CrA was measured in a
small aperture (4$''$) by Roche et al. (1991), who found strong IEB.
Multi--aperture photometry of HD 176386 by Prusti et al. (1994)
indicate extended emission.

\begin{figure}[!ht]
  \begin{center}
    \leavevmode
  \centerline{\epsfig{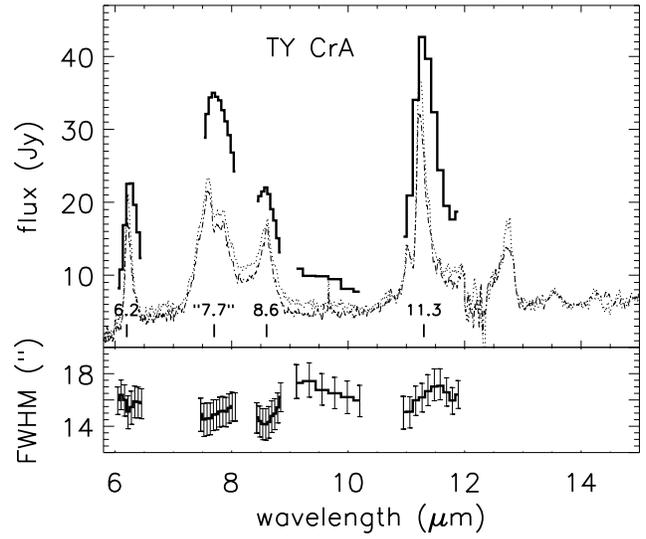}}
  \vspace{-0.5cm}   \end{center}
\caption{Top: Mid infrared spectrum of the spherical component around
TY~CrA.  The CVF scan is shown as solid line and the ISOSWS spectra
of two different epochs are indicated by the dotted and dashed
lines. Bottom: The FWHM as derived from multi--aperture photometry on
the CVF images.}
\label{fig:TYCrA_data_peak}
\end{figure}

In Fig.~\ref{fig:TYCrA_color_113} we show the morphology at
11.28\,$\mu$m.  At about 8$''$ SE from the peak we see a bar like
structure (TY~CrA~bar). The bar is not caused by stray-light or ghost
components, which can be predicted with the optical model of
ISOCAM. The spectrum of the spherical component of TY~CrA is shown in
Fig.~\ref{fig:TYCrA_data_peak}. Photometry was done in half circle
apertures centered on the star at the opposite side of the bar. This
flux was multiplied by two to get an estimate of the total flux
without the bar component. This procedure gives an estimate of the
flux in the spherically symmetric envelope around the star. The
spectrum is rich in IEB and has a very pronounced 8.6\,$\mu$m
band. The 11.05\,$\mu$m feature detected by Roche et al. (1991) is
not resolved by ISOCAM but clearly present in both ISOSWS spectra
(Corporon et al. 1999). The source is very extended ($\sim 2000$\,AU)
in both the IEB and the continuum. Consequently the absolute flux in
both ISOSWS spectra is underestimated by $\sim 40\% $.

\begin{figure}[!ht]
  \begin{center}
    \leavevmode
  \centerline{\epsfig{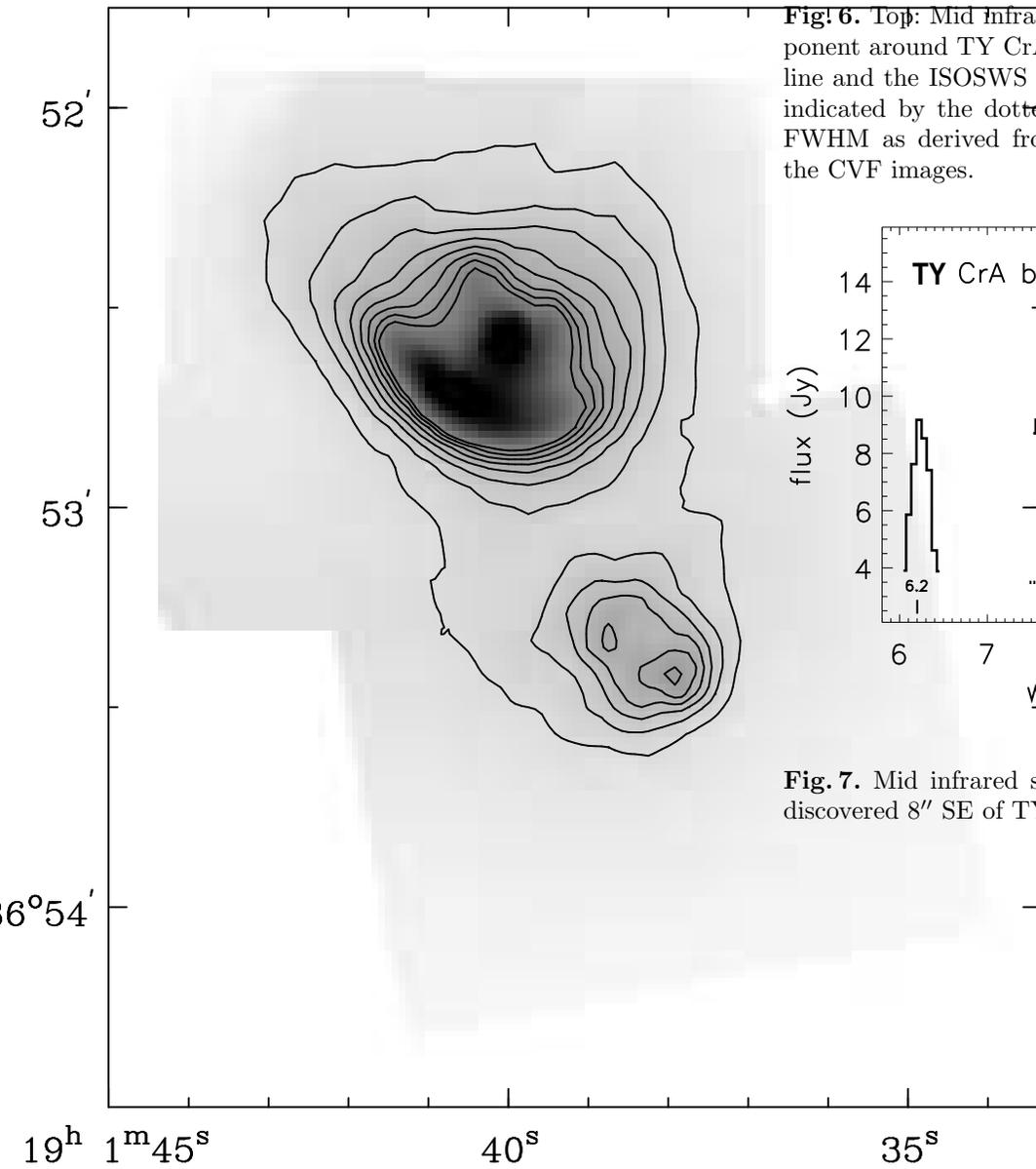}}
  \vspace{-0.5cm}   \end{center}
\caption{Mid infrared spectrum of the bar like structure
discovered 8$''$ SE of TY~CrA.}
\label{fig:TYCrA_data_cmp2}
\end{figure}

In order to deduce the spectrum of the bar like component we have
mirrored the pixels in the half circle used for 41$''$ aperture
photometry and subtracted these images from the original background
cleaned maps. The spectrum of the TY~CrA bar is shown in
Fig.~\ref{fig:TYCrA_data_cmp2}. It shows again strong IEB but weaker
''7.7''and 8.6\,$\mu$m bands as compared to the spherical component of
TY~CrA.

\begin{figure}[!ht]
   \begin{center}
\vspace{-2.0cm}
  \centerline{\epsfig{file=h2191.f5}}
  \vspace{-0.5cm}   \end{center}
\caption{Combined CVF images (J2000) of TY CrA (North) and HD 176386
(South) at 11.28 micron with 3$''$ pixel field of view. Contour levels
are from 0.11 to 0.46 Jy/pixel at intervals of 0.05 Jy/pixel.}
\label{fig:TYCrA_color_113}
\end{figure}

\begin{figure}[!ht]
  \begin{center}
    \leavevmode
  \centerline{\epsfig{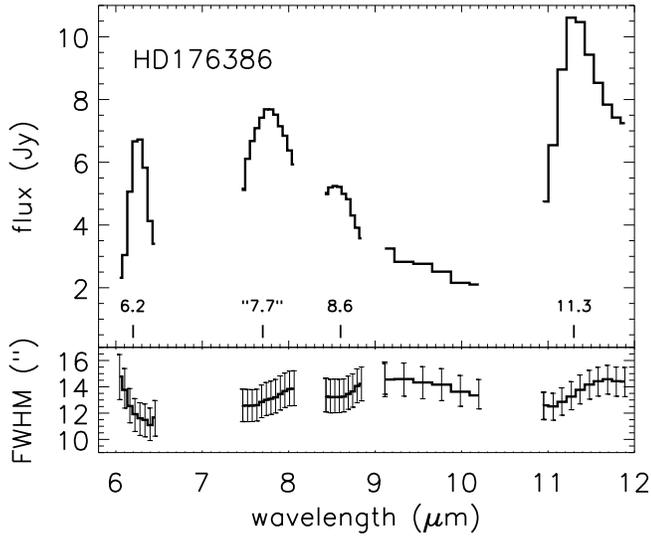}}
  \vspace{-0.5cm}   \end{center}
\caption{Top: Mid infrared spectrum of HD\,176386 main component. Bottom: The FWHM as derived from
multi-aperture photometry on the CVF images.}

\label{fig:HD176386_data}
\end{figure}

\begin{figure}[!ht]
  \begin{center}
    \leavevmode
  \centerline{\epsfig{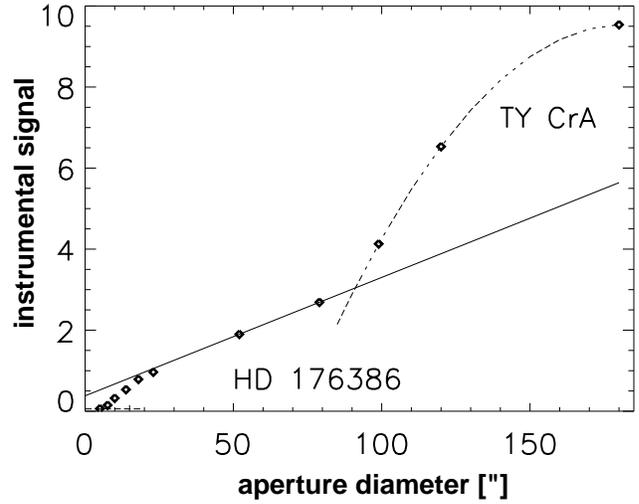}}
  \vspace{-0.5cm}   \end{center}
  \caption{ISOPHT multi--aperture sequence at
    $7.3\,\mu$m on HD 176386. For    apertures larger than 80$''$  
	the adjacent source TY CrA dominates
                        the measured signals (Fig.~\ref{fig:TYCrA_color_113}). }

\label{fig:P04_7p3_hd176386a}
\end{figure}

\begin{figure}[!ht]
  \begin{center} \leavevmode
    \centerline{\epsfig{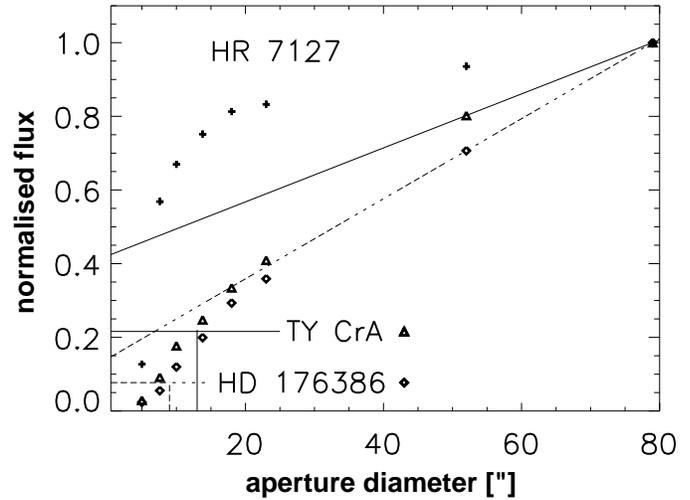}}
    \vspace{-0.5cm} \end{center} \caption{ISOPHT multi--aperture
    sequence, normalised flux at $7.3\,\mu$m on a point source (HR
    7127) and on the extended adjacent sources TY CrA and HD 176386.
    The full and dashed line are for TY~CrA and HD176386,
    respectively. The FWHM is deduced as in Fig.~4 and determined to
    $13 \pm 3''$ (TY CrA) and $9 \pm 3''$ (HD 176386).  }
\label{fig:P04_7p3_hd176386b}
\end{figure}

As in TY CrA, we see two peaks of emission also in HD~176386. The
maximum brightness is at the stellar position. Photometry of the
spherical component was done with half circles in a similar manner as
for TY~CrA. The spectrum is rich in IEB. The band ratios are quite
similar to the bar component of TY~CrA
(Fig.~\ref{fig:HD176386_data}). The FWHM as derived from the ISOCAM
images are indicative of a large extended halo. They are confirmed by
the ISOPHOT multi--aperture sequence at 7.3$\mu$m. In
Fig.~\ref{fig:P04_7p3_hd176386a} we show the ISOPHT measurements of
HD176386.  For apertures above 80$''$ the signals are strongly
contaminated by the adjacent source TY CrA. We derive a source size of
$9 \pm 3''$ (FWHM) for HD176386 and $13 \pm 3''$ (FWHM) for TY CrA
(Fig.~\ref{fig:P04_7p3_hd176386b}). Both size estimates are consistent
with the CVF results.

\begin{figure}[!ht]
  \begin{center}
    \leavevmode
  \centerline{\epsfig{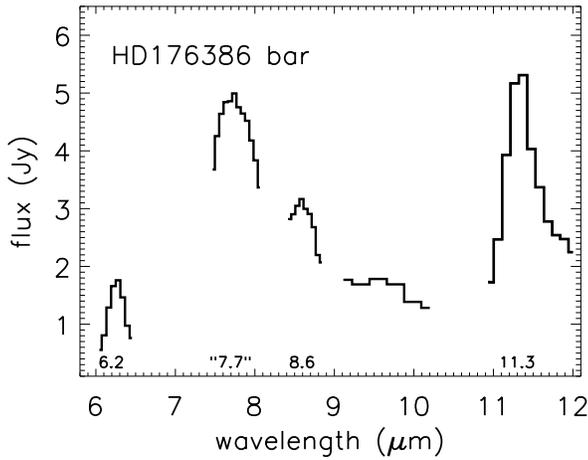}}
  \vspace{-0.5cm}   \end{center}
  \caption{Mid infrared spectrum of the second component of HD\,176386.}
\label{fig:HD176386_data_B}
\end{figure}

In order to derive photometry of the second component of HD176386, we
mirrored the pixels of the half circle 41$''$ aperture and subtracted
the result maps from the original background cleaned images. The
spectrum of the second peak, ''HD176386 bar'', is computed in a 14$''$
aperture. The source is clearly detected at all wavelengths
(Fig.~\ref{fig:HD176386_data_B}). Its spectral shape, in particular of
the 6.2/''7.7'' band ratios, are quite different from the HD176386
main component (Fig.~\ref{fig:HD176386_data}) but similar to TY CrA
main component and HD97048 (Fig.~\ref{fig:HD97048_data}).


\subsection {HD\,179218}
HD\,179218 (also called MWC~614) is a young star of spectral type B9.
It is listed as a Herbig Ae/Be star by Th\'e et al. (1994).  It has no
companion down to a distance of 0.4$''$ (Pirzkal et al. 1997).  To
the best of our knowledge, it has not been observed in the
near-infrared, but has IRAS fluxes in the four bands and shows a 10
$\mu$m emission feature in the low-resolution IRAS spectra.

\begin{figure}[!ht]
  \begin{center}
    \leavevmode
  \centerline{\epsfig{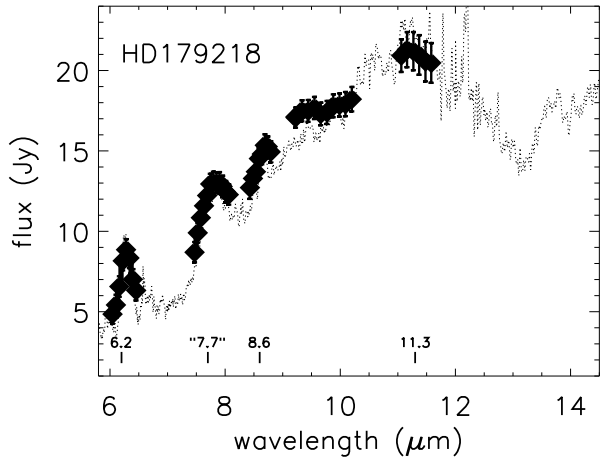}}
  \vspace{-0.5cm}   \end{center}
\caption{Mid infrared spectrum of HD179218. The CVF scan is shown as solid line and the ISOSWS spectra by the dotted line. }
\label{fig:HD179218_data}
\end{figure}

The multi--aperture CVF spectrum is shown in
Fig.~\ref{fig:HD179218_data}. There are pronounced IEB visible at 6.2,
''7.7'' and 8.6\,$\mu$m on top of a strong silicate emission
feature. As discussed by Waelkens et al. (1999) the silicate band of
HD179218 is dominated by substructures due to crystallinity. Those are
absent in spectra of the interstellar medium or in the disks of the
youngest stellar objects. We do not find any clear sign of an extended
structure in the ISOCAM or in the ISOPHT observations. As expected for
a point--like sources, the ISOSWS photometry is in line with our
ISOCAM spectrum.


\subsection{HR\,6000 and HR\,5999}
 
HR\,6000 and HR\,5999 form the proper motion binary system Dunlop 199.
They are located in the Lupus~3 star forming region which is at a
distance of 140$\pm$20\,pc (Hughes et al. 1993). Van den Ancker et
al. (1996) classified HR6000 as a young main sequence object with
spectral type B6. The strong X-ray emission is accounted for by a T
Tauri companion (Zinnecker \& Preibisch 1994). We have a detection of
HR\,6000 above the background level. The star is close to the much
brighter target HR\,5999 so that only the central five pixels could be
used to deduce the spectrum shown in Fig.~\ref{fig:HR6000_data}. By
comparing the spectrum with a black body of 13000\,K, we notice that
it is featureless.


\begin{figure}[!ht]
  \begin{center}
    \leavevmode
  \centerline{\epsfig{file=h2191.f13,width=7.0cm,angle=90}}

  \vspace{-0.5cm}   \end{center}
\caption{Mid infrared spectrum of HR\,6000 together with a black-body
spectrum of 13000\,K (solid line).}
\label{fig:HR6000_data}
\end{figure}

The pre-main sequence star HR\,5999 (also called V856~Sco) has
spectral type A7\,III (Th\'e et al. 1994) and has Rossiter~3930 as
companion (Stecklum et al. 1995), which probably accounts for the
strong X-ray emission.  HR\,5999 has a strong near-IR excess
(Hillenbrand et al. 1992), and has been detected by IRAS at 12, 25 and
60 $\mu$m, as well as at millimeter wavelengths (Henning et al. 1994).
The amount of circumstellar matter associated to the star is of the
order of 0.006 M$_\odot$.  Our mid-infrared spectrum of HR\,5999
(Fig.~\ref{fig:HR5999}) shows clearly a broad silicate emission
feature.  In the CVF spectrum this feature peaks at considerably
shorter wavelengths ($\approx 9.6 \mu$\,m) than observed for HD179218
or Elias ~1 (Hanner et al. 1994). Its intensity is small in
comparison to the underlying continuum, with a ratio of the peak -- to
-- continuum flux of about 1.4.  The FWHM as derived by the ISOCAM CVF
images or the ISOPHT multi--aperture sequence of HR6000 and HR5999 is
typical for point sources.


\begin{figure}[!ht]
  \begin{center}
    \leavevmode
  \centerline{\epsfig{file=h2191.f14,width=7.0cm,angle=90}}
  \vspace{-0.5cm}   \end{center}
\caption{As Fig.~\ref{fig:HD179218_data} for HR\,5999.}
\label{fig:HR5999}
\end{figure}

\subsection {T~Cha}

T~Chamaeleonis is a T~Tauri star of spectral type G8 (Alcal\'a et al.
1993) and the coolest star in our sample. It is superposed on a dark
cloud DC~300.2$-$16.8 between the star forming Chamaeleon~I and II
clouds. Although a priori the later spectral type suggests that it is
physically more difficult to excite transiently heated particles, we
considered this star suitable for our sample as its presumed parent
cloud DC~300.2$-$16.8 shows an unusually high amount of excess
radiation in the mid-infrared (Laureijs et al. 1989).
 
The Hipparcos distance to T~Cha is $66^{+19}_{-12}$\,pc. This is
considerably closer than the estimated distances to the Chamaeleon~I
($160\pm15$) and II ($178\pm18$)\,pc clouds (Whittet et al. 1997). On
the other hand, Terranegra et al. (1999) associate T~Cha to a moving
group of pre-main sequence stars which have the same radial velocity as
the CO gas in DC~300.2$-$16.8. They note, however, that despite sharing
a common proper motion with the group, T~Cha has a deviating radial
velocity measured by Covino et al. (1997). Therefore, unfortunately, it
is not possible to make any firm statements of the possible physical
connection between T~Cha and DC~300.2$-$16.8.

\begin{figure}[!ht]
  \begin{center}
    \leavevmode
  \centerline{\epsfig{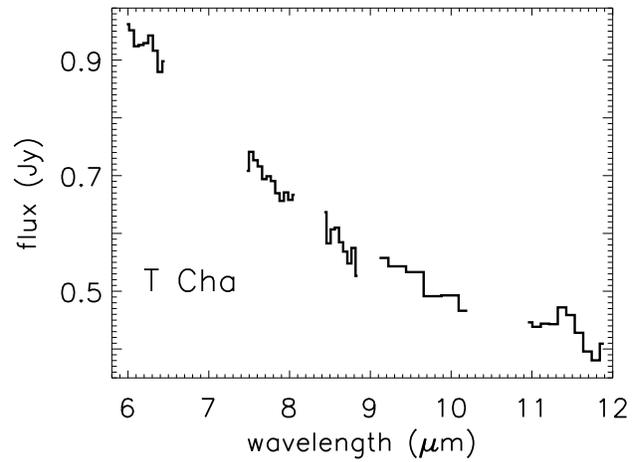}}
  \vspace{-0.5cm}   \end{center}
\caption{Mid infrared spectrum of T~Cha.}
\label{fig:TCHA}
\end{figure}

The spectrum of T Cha is featureless, with no indication of IEB or
silicate emission (Fig.~\ref{fig:TCHA}). The continuum we measure is
about a factor 4--5 larger than the expected photospheric flux (about
0.2 Jy at 6 $\mu$m), possibly due to a circumstellar disk (Alcala et
al. 1993).  Given the lack of dust features it is not a surprising
result that there is no extended mid-infrared emission around
T~Cha. Also the ISOPHT multi--aperture sequence is for apertures below
23$''$ consistent with the point source HR7127. It is possible that a
G8 star cannot excite extended mid-infrared emission. However, given
the fact that the connection to DC300.2-16.8 is uncertain, we cannot
exclude that the lack of extended IEB emission is simply due to the
absence of low-density circumstellar matter associated with the star.

\section {Conclusions}

We find that the mid infrared spectrum of the Herbig Ae/Be stars:
HD~97048, HD~97300, TY~CrA and HD~176386, is dominated by IEB and show
no signature of silicate emission.  All those stars have an extended
halo on scales of up to a few thousand AU.  Two of the stars (TY~CrA
and HD~176386) have a secondary peak of emission at distances of about
1000\,AU from the main component. A secondary peak is found also in
HD97300 and in addition, there is emission in an elliptical ring
structure at a much larger scale ($\sim 7500$\,AU from the star).

The IEB ratios in the secondary peaks show differences when compared
to those of the central regions.  This can be explained as being due
to differences of the excitation of transiently heated particles and
variations of the column density (Siebenmorgen et al. 1998).

The other targets of our sample are at distances similar to the IEB
dominated objects.  They remain point like at the spatial resolution
of ISO ($\sim 5''$ at 6$\mu$m).  Two of those stars (HR6000 and T~Cha)
show featureless mid IR spectra while the other objects (HR5999 and
HD~179218) have a pronounced silicate emission bump.  Because silicate
grains are much larger particles than the IEB carriers (Siebenmorgen
et al. 1992), they are not transiently heated and therefore must be
located close to the star, most likely in a disk component.  This
interpretation is in agreement with the compact emission detected. The
mid IR spectrum of HD~179218 shows both silicate emission and IEB;
nevertheless it is not extended. Consequently the detection of IEB
alone is not a clear indicator to assume an associated extended
emission structure of more than a few hundred AU.

Although the ISOCAM CVF has a known stray-light component we could
confirm the deduced sizes (FWHM) by ISOPHT multi--aperture
photometry. We find that the total flux is in the ISOSWS spectra of
the extended sources systematically underestimated. This is expected
because the ISOSWS flux calibration assumes a point source.  For
point like sources we derive for both ISO instruments a set of
consistent photometric flux density values.

\acknowledgements PIA is a joint development by the ESA Astrophysics
  Division and the ISOPHOT consortium. The ISOCAM data presented in
  this paper was analysed using ''CIA'', a joint development by the
  ESA Astrophysics Division and the ISOCAM Consortium. The ISOCAM
  Consortium is led by the ISOCAM PI, C. Cesarsky.

\end{document}